\documentclass[aps,prl,twocolumn,superscriptaddress,showpacs,floatfix,nofootinbib]{revtex4-1}
\usepackage{graphicx}% Include figure files
\usepackage{dcolumn}% Align table columns on decimal point
\usepackage{bm,palatino,color}
\usepackage[bottom]{footmisc}
\usepackage{mathtools}

\newcommand{\avg}[1]{\langle {#1} \rangle}

\newcommand{\ad}{\hat{a}^{\dagger}}

\newcommand{\ve}[1]{\mathbf{#1}}

\newcommand{\ah}{\hat{a}}

\begin{document}
\preprint{APS/123-QED}

\title{Testing Wavefunction Collapse Models using Parametric Heating of a Trapped Nanosphere}% Force line breaks with \\

\author{Daniel Goldwater}\email{dangoldwater@gmail.com}
%\homepage{dangoldwater@gmail.com}
\affiliation{Department of Physics and Astronomy, University College London, Gower Street, London WC1E 6BT, United Kingdom}

\author{Mauro Paternostro}
\affiliation{Centre for Theoretical Atomic, Molecular, and Optical Physics, School of Mathematics and Physics, Queen's University, Belfast BT7 1NN, United Kingdom}

\author{P. F. Barker}%
%\affiliation{%
\affiliation{Department of Physics and Astronomy, University College London, Gower Street, London WC1E 6BT, United Kingdom}

%\collaboration{CLEO Collaboration}%\noaffiliation%%

\date{\today}% It is always \today, today,

\begin{abstract}
We propose a mechanism for testing the theory of collapse models such as continuous spontaneous localization (CSL) by examining the parametric heating rate of a trapped nanosphere. The random localizations of the center-of-mass for a given particle predicted by the CSL model can be understood as a stochastic force embodying a source of heating for the nanosphere. We show that by utilising a Paul trap to levitate the particle and optical cooling, it is possible to reduce environmental decoherence to such a level that CSL dominates the dynamics and contributes the main source of heating. We show that this approach allows measurements to be made on the timescale of seconds, and that the free parameter $\lambda_{\rm csl}$ which characterises the model ought to be testable to values as low as $10^{-12}$ Hz. 
%\begin{description}
%\end{description}
\end{abstract}

\maketitle
%{\renewcommand*\@makefnmark{}
%\footnotetext{Text}
%\makeatother}
%\let\thefootnote\relax\footnotetext{$^*$dangoldwater@gmail.com}
Dynamical reduction models -- better known as collapse theories -- seek to resolve the measurement problem by inserting a non-linear stochastic term in the Schr\"{o}dinger equation (SE). This would account for genuine collapses of superposition states. In these theories, `localisation' events occur at a frequency scaling with the mass of the system at hand. These are fundamentally different to environmental decoherence~\cite{Joos2003}, and are invoked as the origin of wavefunction collapse. Such modifications to the SE aim to provide a theory capable of describing phenomena at all scales, and are designed to reproduce conventional quantum mechanics (CQM) when dealing with small masses, and  classical mechanics at the macroscopic scale.

One of the most celebrated models of dynamical reduction is the Continuous Spontaneous Localization (CSL) model~\cite{Ghirardi1990}. It is characterised by two parameters:  a length  $r_c$, and a frequency $\lambda_{\rm csl}$. The former provides a length scale above which reduction effects would be relevant, the latter embodies the rate at which spatial superpositions of a single nucleon separated by a distance greater than $r_c$ would collapse. While $r_c$ is generally taken to be $\approx 100$ nm, the value of $\lambda_{\rm csl}$ is the subject of uncertainties~\cite{Fu1977,Bassi2012,Adler2013} and is currently taken to span a range from $10^{-16}$ Hz ~\cite{Ghirardi1990} to $10^{-5}$ Hz ~\cite{Feldmann2012}. A value of $10^{-8\pm 2}$ Hz has been proposed~\cite{Adler2006}, based on the process of image formation on photographic film. The heating rate of ultracold atoms was used to set a value of $10^{-7}$ Hz~\cite{Laloe2014}, while  the maximum allowable heating rate of the intergalactic medium seems to be compatible with $10^{-10\pm 2}$ Hz~\cite{Bassi2010}.

The effects of localization are mathematically very similar to those of decoherence \cite{Bassi2003}, meaning that any experiment built to search for a signature of such a collapse mechanism must minimize the effects of decoherence as much as possible so as to better distinguish the hallmark of the former from that of the latter. However, for a given object, both its rate of localization and decoherence will increase proportionally to mass. To test collapse we must study objects large enough to have an appreciable localization rate, yet small enough that decoherence does not dominate the dynamics. The scale at which this becomes possible is the so-called mesoscopic one, the liminal scale between the well established quantum and classical regimes. Recently, Nimmrichter et al. have shown ~\cite{Nimmrichter2014} that beyond a certain size collapse effects have a sub-linear scaling with size, a result which we corroborate here, and which clearly identifies the regime of interest.

%Recently, various proposals have been made to test collapse theories, and the challenge of reducing decoherence manifests differently depending on the proposal. Many of these proposals do not include detailed analyses of conventional environmental noise, remaining closer to proof-of-principle than fully fledged plans, and as such it is not clear what range of values of $\lambda_{\rm csl}$ they could probe \cite{Sekatski2014,Romero-Isart2011,Emary2014,Arenz2014}. More detailed proposals exist, such as the one put forward in Ref.~\cite{Bateman2014A}, which requires many identical particles to be prepared and detected, and does specify its range of enquiry. Other spectroscopic-like methods \cite{Bahrami2014} are attractive because they in principle only require a lineshape measurement around the mechanical frequency of light scattered by macroscopic oscillator. However, as this lineshape is narrow (order of $\mu$Hz) very long times (in excess of months) would be required for the measurement. A review of approaches to testing collapse models can be found in Ref.~\cite{Arndt2014}.

Lately there have been a considerable number of proposals to test collapse theories, and the challenge of reducing decoherence manifests itself differently depending on the scheme \cite{Sekatski2014,Romero-Isart2011,Emary2014,Arndt2014}. Unfortunately many proposals do not include detailed analyses of conventional environmental noise, and as such, it is not clear what range of values of $\lambda_{\rm csl}$ they could probe. Matter-wave interferometric methods are an attractive means for such tests with a well established range of testable values of $\lambda_{\rm csl}$~\cite{Bateman2014A}. Such settings, however, require the preparation and detection of many identical particles, which makes implementation challenging. Optomechanical proposals \cite{Bahrami2014} are attractive in comparison, because they in principle require only a line shape measurement of the light scattered by an intra-cavity macroscopic oscillator, and do not require ground state cooling. However, as such line shape would be narrow (order of $\mu$Hz) very long times (in excess of months) would be required for the measurement. A recent proposal has been made based on dynamical decoupling \cite{Arenz2014}, which is promising but may also be constrained by long testing times. 

The concept of utilising the energy gain of a harmonic oscillator to test CSL was first suggested by Adler~\cite{Adler2004} and more recently re-examined in Refs.~\cite{Bera2015,Diosi2014}. Whilst this method is relatively straightforward and therefore attractive for implementation, the ability to test for CSL is very dependent on a detailed and realistic inclusion of conventional decoherence mechanisms. In order for the {\it classical} approach put forward, for instance, in Ref.~\cite{Diosi2014} to be effective, the collapse mechanism must be able to induce the excitation of a rather substantial number of thermal phonons in a given oscillator. This is, in general, not the case for a large range of values of the parameters that characterize a given collapse model and such approaches can only assess the largest of their conjectured values.

In this paper, we propose an experimentally viable way to explore CSL on the mesoscale by utilising a cavity-cooled, single-charged nanosphere trapped in a Paul trap \cite{Millen2015}. Measurements can be made with a single trapped particle in less than 100 seconds and, under optimal conditions, we find this scheme capable of probing $\lambda_{\rm csl}$ to values as low as $10^{-12}$ Hz, thus going significantly beyond the literature reported so far. Most importantly, our protocol allows for the discrimination between collapse effects and mis-characterised conventional noise ones-- a challenge which has to our knowledge remained unaddressed so far.
% We have modelled all relevant noise sources affecting the system and show that not only does CSL contribute a heating rate for a harmonic oscillator, but that given the right parameters this heating rate appreciably dominates the dynamics.  

\begin{figure}
\centering
\includegraphics[width=.4\textwidth]{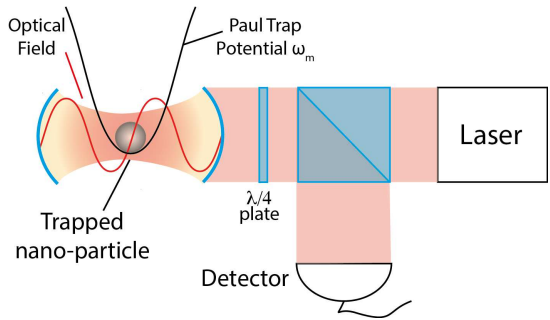}
\caption{Schematic of the experiment, in which the particle is levitated by the electric field of the Paul trap, and cooled by the optical cavity. }
\label{figure.NanoParticleSchematic2}
\end{figure}

We explore the possibility for an optomechanical test of a form of CSL described in Ref.~\cite{Nimmrichter2014,Bahrami2014,Bassi2003}, in which the effects of spontaneous localization are modelled as a delta-correlated stochastic noise source $w_t$. % possessing the autocorrelation function $\avg{w_t(t)w_t(t')}=\delta(t-t')$. 
This approach is valid when the scale of spatial-superposition separations is less than $r_c$. The noise term $w_t$ will occur in the dynamical equations of the system as an extra Langevin force~\cite{QuantumNoise, Bahrami2014}. Its effect on the dynamics of a mechanical oscillator would depend on the size and density of the object collapsing, and the two parameters $r_{\rm c}$ and $\lambda_{\rm csl}$ characterizing the model. Conveniently, we can represent the effects of the localization process via a diffusion operator characterised by the coefficient ${D}_{\rm csl}$ and appearing in the master equation describing the particle in the same way as conventional heating sources. The diffusion coefficient takes the form~\cite{Nimmrichter2014} 
\begin{equation}
\label{csldiff}
{D}_{\rm csl}=\frac{\hbar}{m \omega_m} \frac{\lambda_{\rm csl}}{r_{c}^2}\alpha,
\end{equation}
where $\alpha$ is a geometry-dependent factor, which for a sphere is given by 
\[
\alpha=\left(\frac{m}{m_0}\right)^2\left[e^{-R^2/r_c^2}-1+\frac{R^2}{2 r_c^2}(e^{-R^2/r_c^2}+1)\right]\frac{6 r_c^6}{R^6}.
\]
%\begin{equation}
%$\alpha={16 \pi^2 \rho^2 r_c^4}R^2/({3 m_0})$.
%\end{equation}
Here $m_0 = 1$ amu and  $R$ is the sphere's radius. %This will form an effective heating rate, which we will derive through solving the master eqation. 

\noindent
{\it The Protocol.--} We now describe the scheme that we propose to test the CSL model. We levitate a charged nanosphere in a hybrid trap consisting of a Paul trap and an optical cavity, and use them to cool its motion to a temperature corresponding to a low occupation number. We then turn off the optical field (and hence the cooling) and let the dynamics evolve for a certain amount of time before measuring the energy of the oscillator again. A model including the effects of CSL predicts it will have heated more than one would expect due to conventional noise sources alone. If the measured energy matches that predicted by conventional noise sources, we will have provided evidence against CSL to within a certain range of $\lambda_{\rm csl}$, whereas a higher measured energy would indicate some other dynamics at play, in favour of collapse theories.  

We divide the procedure into two phases: a cooling phase, and free evolution. For the purpose of testing CSL, it is the second phase that is important. In this period of free evolution the nanosphere is levitated using a single electric potential, which could be provided by a number of generic trap architectures. The mechanism of cavity cooling for nanoscale objects is well established~\cite{Paternostro2006,Millen2015,Wilson-Rae2007,Geiseler2012,Romero-Isart2011,Genes2007,Barker2012,Barker2010,Chang2010,Romero-Isart2010,Pflanzer2012,Romero-Isart2011b,Monteiro2013}, and relies on having the particle sit in two potential wells: one of which traps, and one (or both) of which cool. Though these potentials are traditionally provided by an optical cavity populated with two distinct optical fields,  Ref.~\cite{Millen2015,Fonseca2015} shows that cooling is possible using a Paul trap in conjunction with a single-mode optical cavity. While both these systems are required for the cooling phase, the Paul trap alone suffices to levitate it.

This is appealing, because in the low pressure scenario of a particle levitated solely by optical fields the dominant source of heating is the scattering of cavity photons~\cite{Chang2010}. %As stated in the introduction, the success of our proposal depends upon the visibility of the CSL induced heating over and above conventional heating sources. 
By using a hybrid trap we can cool the particle to a desired temperature and then turn off the optical field completely, leaving the particle suspended in the Paul trap alone. This ability to turn off the optical field without `losing' the particle means that we can do away with what would otherwise be the dominant cause of heating -- optical scattering -- and thus greatly reduce the conventional heating sources that would otherwise mask the CSL effects. 

We emphasize that although we require cooling, we do not need to achieve the ground-state energy. Indeed, the simple comparison between the initial phonon number $n_0$ and the final one $n_f$ after the period of free evolution will give us information on the heating rate ~\cite{Supplementary}.
%Likewise, a discussion of the bulk temperature of the sphere (which is an important factor, as it determines the rate of decoherence due to blackbody radiation, and is related to the cooling phase in that it is chiefly determined by the intensity of the laser), will be left to the SI.  

The period of free evolution is governed by the Hamiltonian $\hat H=\hat H_0+\hat H'$, where $\hat H_0=\hbar \omega_m \ad \ah$,  $\omega_m$ is the secular frequency of the Paul trap, $\ad, \ah$ are the creation and annihilation operators for the centre of mass motion of the sphere respectively, and $H'$ represents the interaction between system and environment. We can then solve the master equation $\dot{\rho}=-({i}/{\hbar})[\hat H,\rho]$ for the oscillator~\cite{Rodenburg2015}. The forms of coupling to noise sources in $H'$ determine their effects on the master equation \cite{Romero-Isart2011b}. We have explored each noise source in detail, examining collisions with the background gas, blackbody radiation, acoustic noise affecting the trap, Johnson and patch potential noise from the electrodes, micromotion from the trap's driving frequency, and anisotropy of the sphere~\cite{Supplementary}. We group these noise sources as momentum diffusion, occurring at rate $D_{\rm diff}$, position diffusion at rate $D_{\rm pos}$, and momentum dissipation at rate $\Gamma$, thus getting the dynamical equation ~\cite{Rodenburg2015}
\begin{equation}
\label{eq.MasterGeneric}
%\begin{aligned}
\dot{\rho}=-\frac{i}{\hbar}[\hat H_0,\rho]-\sum^2_{j=1}D_{j}[\hat X_j,[\hat X_j,\rho]]-\Gamma [Q_z, \{ \hat P_z,\rho\}]
%\dot{\rho}=&-\frac{i}{\hbar}[\hat H_0,\rho]-D_{\rm diff}[\hat Q_z,[\hat Q_z,\rho]]\\
%&-D_{\rm pos}[\hat P_z,[\hat P_z,\rho]]-\Gamma \{ \hat P_z,\rho\}
%\end{aligned}
\end{equation}
where $\hat {\bm X}=(\hat Q_z,\hat P_z)$ is the vector of quadratures of the nanosphere $\hat Q_z=\ah+\ad$ and $\hat P_z=i(\ad-\ah)$, and ${\bm D}=(D_{\rm diff}, D_{\rm pos})$. Our analysis of the various noise sources, including the possible effects of anisotropy of the nanosphere (and the consequent non-uniform distribution of the charge) finds all but gas collision and blackbody radiation to be negligible~\cite{Supplementary}, giving us 
%\begin{equation}
%\label{eq.diffterms}
$\Gamma=(\gamma_{\rm gas}+\gamma_{\rm {bb,e}}+\gamma_{\rm {bb,a}})/4$ and $D_{\rm diff}=D_{\rm gas}+D_{\rm bb}+D_{\rm csl}$,
%\end{equation}
where $D_{\rm csl}$ is given in Eq.~\eqref{csldiff} and
\begin{equation}
D_{\rm gas}=\frac{\gamma_g k_B T_{\rm env}}{2 \hbar \omega_m},~~D_{\rm bb}=\frac{k_B(\gamma_{\rm {bb,e}} T_{\rm int}+\gamma_{\rm {bb,a}} T_{\rm env})}{2 \hbar \omega_m}.%\\
%D_{csl}&=\frac{\hbar}{m \omega_m}\frac{\lambda_{csl}}{r_c^2}\alpha
\end{equation}
Here, $\gamma_{\rm gas},\gamma_{\rm {bb,e}},\gamma_{\rm {bb,a}}$ are the damping constants related to gas collisions, blackbody emission and blackbody absorption respectively. The environmental temperature is $T_{\rm env}$, $\omega_m$ and $m$ is the mass of the nanosphere.

\noindent
{\it Heating Rate of a Trapped Particle.--}
The mean occupation number (phonon number) of the trapped particle as a function of time $\avg{n}_t$ is determined via Eq.~(\ref{eq.MasterGeneric}).
 Assuming a thermal state~\cite{Wilson-Rae2007,Marquardt2007,Romero-Isart2011b}, the expression for the rate of change of the average phonon number simplifies to 
$\avg{\dot{n}(t)}=-\Gamma \avg{n(t)}+D_{\rm diff}, $
which has the solution
\begin{equation}
\avg{n(t)}=e^{-\Gamma t}\left(n_0-\frac{D_{\rm diff}}{\Gamma}\right)+\frac{D_{\rm diff}}{\Gamma}, \label{eq.HeatingRate}
\end{equation}
where $n_0$ is the initial average number. In Fig.~\ref{fig.1sHeatingCompare} we plot the expected mean phonon number over the first second of free evolution, starting from an initial number of $n_0=50$, showing the heating when the CSL mechanism is included (solid blue line), and when it is not (dashed orange line).
%\begin{figure}[!ht]
%  \centering
%  \subfloat[][Including CSL]{\includegraphics[width=.25\textwidth]{./images/CSLHeating1s.pdf}\label{fig.1sHeatingCompareCSL}}
%  \subfloat[][CQM]{\includegraphics[width=.25\textwidth]{./images/CQMHeating1s.pdf}\label{fig.1sHeatingCompareCQM}}
%   \caption{Mean phonon number $\avg{N}$ over one second according to CSL and CQM. Simulations produced using $R=100$nm, pressure =$10^{-9}$mbar, $\omega_m=10$kHz, $\rho=2300$kg/m$^3$, environment temperature = 4K, and $\lambda_{csl}=10^{-8}$ and 0 for the two plots respectively. The CSL case heats significantly more than the CQM.}
%   \label{fig.1sHeatingCompare}
%\end{figure}  
%The mean phonon number for the thermal state of the oscillator evolves over one second for each theory from an initial number $n_0$. 
In our case, the diffusive terms in Eq.~(\ref{eq.HeatingRate}) dominate over the dissipative, resulting in an expression that is approximately linear, as seen in Fig.~\ref{fig.1sHeatingCompare}. The inclusion of the CSL mechanism results in a heating rate of the nanosphere motion of about 2500 phonons/s, which is in stark contrast with the $\lambda_{\rm csl}=0$ case, where only $\sim 350$ phonons/s are achieved for the parameters used in our simulations.

\begin{figure}[b]
\centering 
\includegraphics[width=.4\textwidth]{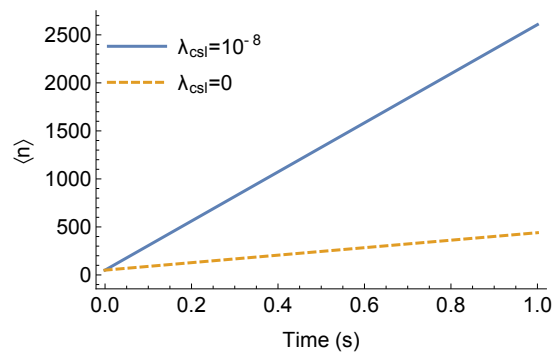}
\caption{{\bf(Color online)} Expected phonon number $\avg{n(t)}$ over one second with and without collapse effects. We have used $n_0=50$, $R=100$ nm, a pressure of$10^{-12}$ mbar, $\omega_m=5$ kHz, $\rho=2300$ kg/m$^3$, $T_{\rm env}= 4$ K, and an internal temperature of 65 K. The solid (dashed) line is for a CSL mechanism characterized by $\lambda_{\rm csl}=10^{-8}, (0)$ Hz, i.e. the Adler value (no CSL mechanism).}
\label{fig.1sHeatingCompare}
\end{figure}

\noindent
{\it Differentiating CSL from Conventional Noise.--} If a final occupation number is recorded that  agrees with the model of $\lambda_{\rm csl}=0$, we interpret it as falsifying a certain range of $\lambda_{\rm csl}$. However, if we measure a higher phonon number, we can infer that some extra heating process is present, possibly collapse effects. However, an objection could be made that the increased heating would more likely result from mis-characterising the environmental noise sources present.  

Therefore, an important requirement is how to correctly identify CSL, and distinguish it from other noise sources. Indeed, this problem is generic to any test of collapse theories. We can address this problem by monitoring the conventional noise sources, such as blackbody radiation and gas collisions, by varying the associated parameters and determining the effect they have on the heating rate. For example, if the heating is strongly dominated by CSL, then varying the pressure will have little effect over some range, while a gas-collision dominated process results in an almost linear relation between pressure and heating rate.
\begin{figure}[b]
%\centering
%\captionsetup[subfigure]{position=top, labelfont=bf,textfont=normalfont,singlelinecheck=off,justification=raggedright}
%\subfloat[][%]
{\bf (a)}\hskip3cm{\bf (b)}\\
{\includegraphics[width=.22\textwidth]{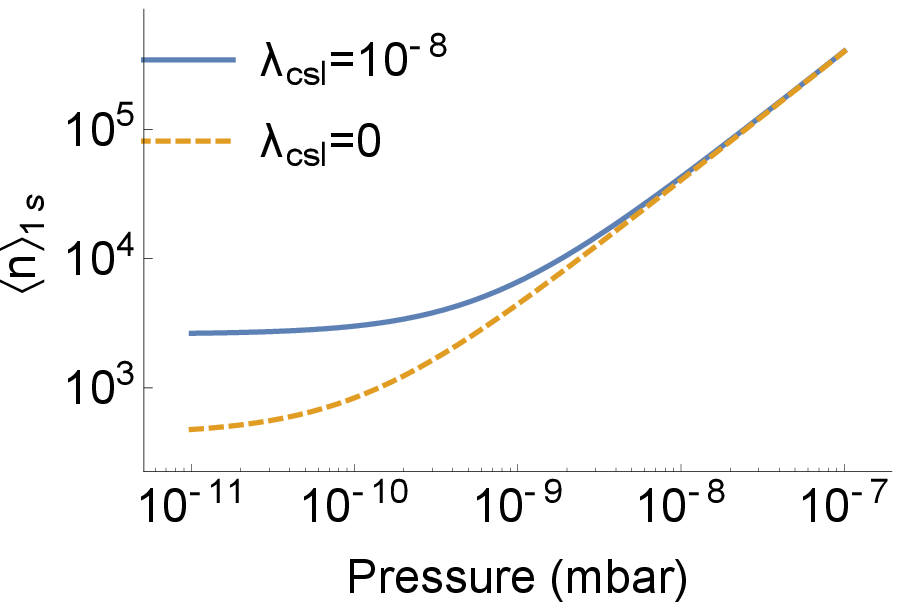}\label{graph.1sLogPressureCompare.eps}}\quad
%\subfloat[][%]
{\includegraphics[width=.22\textwidth]{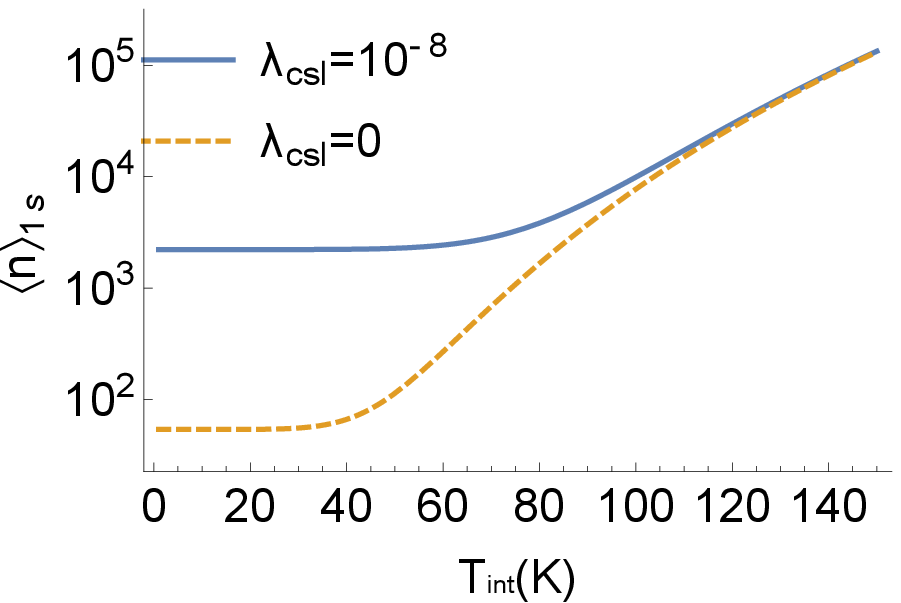}\label{graph.1sLogTintCompare.eps}}\\
%\subfloat[][%]
{\bf (c)}\hskip3cm{\bf (d)}\\
{\includegraphics[width=.22\textwidth]{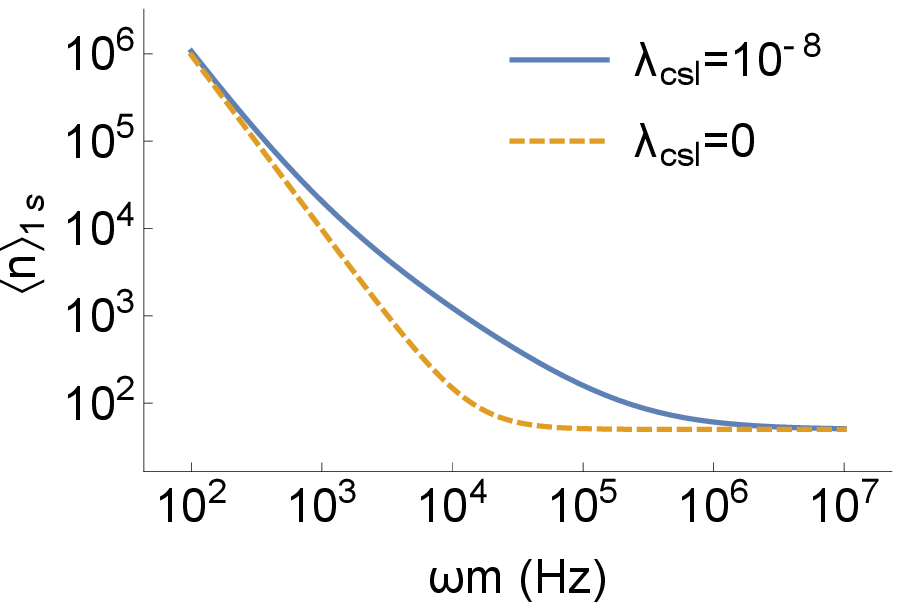}\label{graph.1sLogFrequencyCompare.eps}}\quad
%\subfloat[][%Comparison of Pressure]
{\includegraphics[width=.22\textwidth]{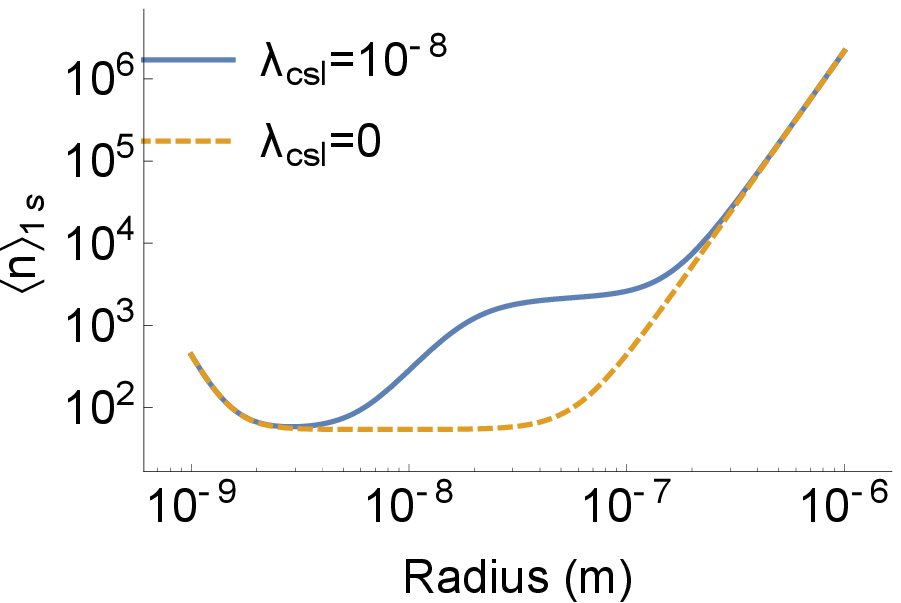}\label{graph.1sLogRadiusCompare.eps}}
\caption{{\bf(Color online)} Phonon number after 1s as predicted by with (blue line) and without (dashed orange line) collapse effects as we vary the pressure in panel {\bf (a)}, bulk temperature in panel {\bf (b)}, mechanical frequency in panel {\bf (c)}, and radius of the sphere in panel {\bf (d)}.
%\ref{graph.1sLogTintCompare.pdf}. 
Except where a parameter is under investigation, we have used the same values as in Fig.~\ref{fig.1sHeatingCompare}.}\label{fig.LogCompare}
\end{figure}

Fig.~\ref{fig.LogCompare} summarises such behaviours, and displays the trend followed by the mean phonon number after one second when varying different parameters for the cases of $\lambda_{\rm csl}= 10^{-8}$ Hz and for $\lambda_{\rm csl}=0$.  In Fig.~\ref{fig.LogCompare} {\bf (a)}, we examine the effect of varying the background pressure when we include or exclude CSL. The difference in response between the theories is instructive. Without CSL the effect of an increasing pressure can be seen across the whole range, whereas for CSL there is a region of immunity where CSL dominates the dynamics. Likewise for the internal temperature depicted in Fig.~\ref{fig.LogCompare} {\bf (b)}: the point at which this significantly influences the heating is different for the two theories. The response to a varying mechanical frequency also takes a different shape, as seen in Fig.~\ref{fig.LogCompare} {\bf (c)}. Most interesting is the effect of a varying radius shown in Fig.~\ref{fig.LogCompare} {\bf (d)}. Our findings agree with those of Ref.~\cite{Nimmrichter2014}: objects must be large enough to have an appreciable collapse rate, but small enough that decoherence does not dominate the dynamics in order for us to observe collapse effects. Such request is met for $R\in[10 , 100] {\rm nm}$, roughly. For all of these we have compared the Adler value for $\lambda_{\rm csl} = 10^{-8}$ Hz with the case of no collapse effects. However the same differences persist for any chosen non-zero value of $\lambda_{\rm csl}$, though being more pronounced for higher collapse rates.

%Each of these parameters has a region where the two theories diverge most. The location, shape and range of the optimal region for a given parameter depends in turn upon the other factors affecting the dynamics of the nanosphere. 
In order to find the best conditions for testing CSL it is necessary to numerically optimise all the parameters simultaneously. In Fig.~\ref{graph.OptimisedLimitPlot}, we show the range of $\lambda_{\rm csl}$ which can be probed for an illustrative set of experimental conditions. We see that, as one would expect, the testable range depends ultimately upon the conditions that can be achieved, most relevantly the minimum value of both environmental pressure and internal temperature. The value of the initial phonon number and evolution time also play a significant role: after enough time, the phonon ratio will tend to the heating rate ratio as
${\avg{n(\infty)}_{\rm csl}}/{\avg{n(\infty)}_{\rm cqm}}\to{\avg{\dot{n}}_{\rm csl}}/{\avg{\dot{n}}_{\rm cqm}}$,
and the time required to approximate this depends on the initial phonon number and heating rates. A lower initial phonon number or longer evolution times would promote each plot in Fig.~\ref{graph.OptimisedLimitPlot}, as a given set of parameters would be capable of probing a lower value of $\lambda_{\rm csl}$.

\begin{figure}
\centering
\includegraphics[width=.4\textwidth]{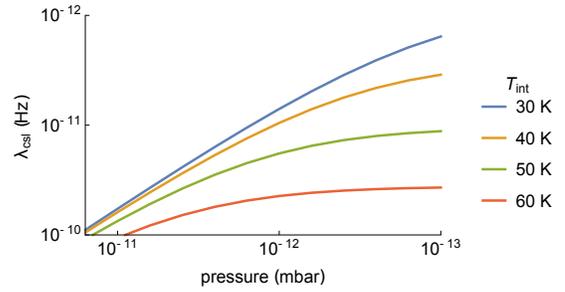}
\caption{{\bf(Color online)} Testable range of $\lambda_{\rm csl}$ as a function of pressure, for a set of  internal temperatures. The pressure shown is where $\avg{n(100{\rm s})}_{\rm csl}/\avg{n(100 {\rm s})}_{\rm cqm}\geq 1.2$ after 100 s with initial phonon number of $n_0=50$. The sphere radius and mechanical frequency have been optimised for each individual data point's temperature and pressure. }
\label{graph.OptimisedLimitPlot}
\end{figure}

\noindent
{\it Experimental Feasibility. --}
Charged silica particles of 200 nm have been trapped and cooled to milliKelvin temperatures in a Paul trap using cavity cooling to milliKelvin temperatures ~\cite{Fonseca2015} thus demonstrating the key experimental components required for our proposal.  Pressures down to the $10^{-11}$ mbar range and internal particle temperatures in the 10 K range can be obtained for this setup using standard cryopumps. An important component of these experiments is the measurement of the oscillator energy following parametric heating by CSL and conventional noise. These can be carried out by performing a homodyne measurement of position as a function of time which allows the determination of mean occupation number. Importantly, the light-cavity detuning can be non-perturbative during the measurement such that it neither cools nor heats the particle during this time. This ability to control the cooling rate also allows us to avoid giving strong kicks to the particle when we turn off the optical field after the particle is cooled to the desired energy.

\noindent
{\it Concluding Remarks.--}
We have shown that the parametric heating rate of a trapped nanosphere provides a viable mean of testing CSL. Central to the success of this scheme is the minimization of all sources of environmental decoherence. The two-stage `cool and release' protocol that we have illustrated allows us to exploit optical cooling while avoiding problematic scattering noise that would otherwise dominate the dynamics. We remark that, owing to established techniques and common experimental settings, the experiment can be performed rapidly, and is repeatable upon a given nanosphere. Remarkably, the central experimental set-up has already been demonstrated~\cite{Millen2015,Fonseca2015}. Our scheme can be used to distinguish CSL from other noise sources -- an essential condition for inferring the existence of collapse effects from an experiment. Further, we have shown that the parameter range of CSL that is testable using our proposal is broad, and can readily be expected to probe  $\lambda_{\rm csl}= 10^{10}$ ($\lambda_{\rm csl}=10^{-12}$ ) using a background pressure of $10^{-11}$ mbar ($10^{-13} $ mbar) and an internal temperature of 60 K (20 K). Based on state of the art, values as low as $\lambda_{\rm csl} =10^{-8}$ could be tested imminently, and  $\lambda_{\rm csl}=10^{-12}$  plausibly in the next few years. 

%Again, we emphasise that the range of values probable is primarily constrained by four factors: the background gas pressure, the internal temperature of the nanosphere, the duration of the free evolution period, and the threshold minimum ratio of we require between the predicted occupation numbers with and without CSL. If the pressure or internal temperature can be lowered further, the free evolution extended, or the detectability threshold relaxed, then the protocol can push even further into the parameter range of $\lambda_{\rm csl}$. 

\noindent
{\it Acknowledgements.--}MP acknowledges the John Templeton Foundation (grant ID 43467),  the UK EPSRC (EP/M003019/1), and the EU FP7 grant TherMiQ (Grant Agreement 618074) for financial support. DG and PFB also acknowledge the UK's EPSRC (EP/H050434/1, EP/J014664/1, EP/K026267/1), and thank H. Ulbricht and M. Bahrami for useful discussions, and Jonathan Underwood for his help in modelling the rotational dynamics of the sphere.

%\bibliography{library.bib,ManualLibrary.bib}

%%%%%%%%%% Merge with supplemental materials %%%%%%%%%%
\pagebreak
\widetext
\begin{center}
\textbf{\large Testing wavefunction collapse models using parametric heating of a trapped nanosphere -- Supplementary Information}
\end{center}
%%%%%%%%%% Merge with supplemental materials %%%%%%%%%%
%%%%%%%%%% Prefix a "S" to all equations, figures, tables and reset the counter %%%%%%%%%%
\setcounter{equation}{0}
\setcounter{figure}{0}
\setcounter{table}{0}
\setcounter{page}{1}
\makeatletter
\renewcommand{\theequation}{S\arabic{equation}}
\renewcommand{\thefigure}{S\arabic{figure}}
\renewcommand{\bibnumfmt}[1]{[S#1]}
\renewcommand{\citenumfont}[1]{S#1}
%%%%%%%%%% Prefix a "S" to all equations, figures, tables and reset the counter %%%%%%%%%%

\title{Testing wavefunction collapse models using parametric heating of a trapped nanosphere -- Supplementary Information}% Force line breaks with \\

\author{Daniel Goldwater}
\homepage{dangoldwater@gmail.com}
\affiliation{Department of Physics and Astronomy, University College London, Gower Street, London WC1E 6BT, United Kingdom}

\author{Mauro Paternostro}
\affiliation{Centre for Theoretical Atomic, Molecular, and Optical Physics, School of Mathematics and Physics, Queen's University, Belfast BT7 1NN, United Kingdom}

\author{Peter F. Barker}%
%\affiliation{%
\affiliation{Department of Physics and Astronomy, University College London, Gower Street, London WC1E 6BT, United Kingdom}

\maketitle
\section{Cooling the Sphere}

Ref.~\cite{Millen2015} give a comprehensive mechanism for cooling the motion of a sphere in a hybrid type trap. Though they achieve an experimental temperature of 10 K, improved to millikelvin in ~\cite{Fonseca2015}, the same technique should in principle be able to achieve ground state cooling and is limited primarily by the finnesse of the mirrors. When translating their results to our scenario however, we must be a little careful. Along the axis of interest, there are three motional frequencies to be considered in the hybrid scheme: $\omega_d$, which is the frequency of the AC voltage applied to the electrodes to keep the trap stable; $\omega_s$ (or secular frequency), which describes the oscillation of the sphere in the electric potential;  $\omega_c$, the frequency of the sphere's motion inside a well of the cavity potential. For our period of free evolution, we are considering the motion inside the electric potential $\omega_s=\omega_m$, whereas the goal of Ref.~\cite{Millen2015} was to cool $\omega_c$. The steady state phonon number  achievable through cooling is given by~\cite{Romero-Isart2011b,Pflanzer2012}
\begin{equation}
N_{\rm ss}= \left( \frac{\kappa+\kappa_{\rm sc}}{4 \omega_c}\right)^2+\frac{\Gamma_{\rm sc}+\Gamma_{\rm others}}{\Gamma_-}
\end{equation}\label{eq.NsteadyState} 
 where we have used $N$ to refer to the phonon number during the cooling period and $n$ for the phonon number during free evolution. $\Gamma_{others}$ refers to  heating processes other than optical scattering, which are all negligible compare to $\Gamma_{sc}$, and $\kappa_{sc}$ is the rate of photon loss from the cavity due to scattering, and the time averaged cooling rate is given by~\cite{Millen2015}
\begin{equation}
\Gamma_-=g^2 k\left[\frac{1}{(\Delta-\omega_c)^2+\frac{\kappa^2}{4}}-\frac{1}{(\Delta+\omega_c)^2+\frac{\kappa^2}{4}} \right]
\end{equation}
in which $k$ is the wavenumber of the laser, and the optomechanical coupling is given by
\begin{equation}
g^2= \frac{1-J_0(4 k X_d)}{2 m \omega_c} \hbar k^2 a_c^2 \left(\frac{3 V}{2 V_c}\frac{\epsilon_r-1}{\epsilon+2}\omega_l\right)^2
\end{equation}
where $a_c^2$ is the number of photons, of frequency $\omega_l$ circulating in the cavity, whose mode volume is $V_c$. Here, $\epsilon_r$ is the relative dielectric permittivity, $J_0$ is the zeroth Bessel function of the first kind and $X_d$ is the amplitude of the AC drive voltage. For a finnesse of $10^5$, ground state cooling within the optical potential should be achievable using an input laser power of $10^{-4}$ W. Using $\avg{N}(t)=\avg{N}_{\rm therm}e^{-\Gamma_- t}+N_{\rm ss}\left( 1-e^{\Gamma_- t}\right)$ where $\avg{N}_{\rm therm}$ is the mean phonon number the oscillator would have if thermalised with its environment, we can see that $\avg{N}(t)\approx\avg{N}_{\rm ss}$ within about a microsecond, so there is no time impediment to achieving the cold states we require. 

Cooling to a given phonon number in the optical potential does not translate directly into that phonon number being occupied in the Paul trap once the cavity field is turned off, due to the different trap frequencies. Supposing the antinodes of the optical well and Paul trap are perfectly aligned, when we turn the optical potential off the sphere should have $n_0=N_{\rm ss}{\omega_c}/{\omega_s}$, where $n_0$ is the phonon number in the Paul trap immediately after the cavity is switched off and and $N_{ss}$ is the phonon number in the optical trap immediately before. However, if we take into account some displacement between the two trap centers $\delta_x$, then we would expect the that $n_0 \geq N_{\rm ss}{\omega_c}/{\omega_s}+m \omega_s \delta_x^2/(2 \hbar)$, where the inequality accounts for the fact that it is also possible for the optical field to impart momentum to the sphere as it is turned off. The ability for this displacement to impart phonons requires that we ensure $\delta_x \lesssim 0.5$nm. 

When we consider these factors, it becomes clear that our predictive knowledge of the phonon number at the beginning of the free evolution $n_0$ is far from perfect. This is important of course, because it is by comparing $n_0$ with $n_f$ that we hope to learn anything. However, it is possible for us to gain experimental, not just predictive knowledge of $n_0$. A `dry run' is possible, in which rather than turning the cavity field completely off we reduce it to such a low power that it no longer traps the particle, but does remain coupled to the particle's position. As such we can use it to measure $n_0$ via the techniques laid out in Ref.~\cite{Paternostro2006}. This information can be used heuristically to better align trap centres, and also to build up a statistical picture of $n_0$, providing a benchmark for the actual experiment in which the cavity field is turned off completely.

\section{The CSL mechanism as a source of noise}

The premise of the experiment is that since collapse effects can be thought of as physically, if not ontologically, equivalent to those of decoherence\footnote{By this we mean that the effects of localisations are, at the level of the density matrix, indistinguishable from those of decoherence. Collapse will destroy coherences and localise a wavefunction. However, the \emph{meaning} of this is profoundly different to usual: the system is being driven into a series of definitive states according to certain probabilities, it is not just being reduced to superposition states who have lost their coherence, and appear to be definite.}, a search for collapse is in some ways equivalent to a search for a new source of decoherence. More specifically, in our case we are searching for a Brownian noise source which heats a harmonic oscillator.  Here we demonstrate that frequent random localisations can be modelled as Brownian noise, drawing on Refs.~\cite{Nimmrichter2014, Bahrami2014}. The modified master equation is given by $\dot{\rho}=(\mathcal{L}+\mathcal{L}_{\rm csl})\rho$, where
\begin{equation}
%\begin{aligned}
\mathcal{L}_{\rm csl}\rho%=\frac{-D_{\rm csl}}{\hbar}[\hat{x},[\hat{x},\hat{\rho}]]\\
=\frac{\lambda_{\rm csl}}{\pi^{3/2}r_{\rm csl}^3m_0^2}\int d^3 x\left[M(\ve{x})\rho M(\ve{x})-\frac{1}{2}\{\rho,m^2(\ve{x}) \} \right]
%\end{aligned}
\end{equation} 
and $M(\ve{x})=\sum_n m_n\exp[{-\frac{(\ve{x}-\ve{r}_n)^2}{2r^2_{\rm csl}}}]$. Here $\ve{r}_n=\ve{r}+\ve{r}_n^0+\Delta \ve{r}_n$ stands for the relative coordinate of particle $n$ in relation to the centre of mass (C.O.M) coordinate $\ve{r}$, with $\ve{r}_n^0$ and $\Delta \ve{r}_n$ describing the motion of the particles bound within the solid. $M(\ve{x})$ is a Gaussian averaged mass density function of the $N$ particle system. As shown in Sec. 8.2 of Ref.~\cite{Bassi2003}, by taking the range of motion for each constituent particle to be much smaller than $r_c$, it is possible to effectively separate the internal dynamics from those affecting the centre-of-mass (COM), meaning that we can re-write the CSL master equation in terms of the C.O.M alone, neglecting the internal dynamics. We thus have
\begin{equation}\label{eq.reducedCSLmaster}
\mathcal{L}_{\rm csl}\rho=\frac{r^3_c \lambda_{\rm csl}}{\pi^{3/2}m_0^2}\int d^3ke^{-r^2_ck^2}|\varrho(\ve{k})|^2(e^{i\ve{k}\cdot \ve{r}}\rho e^{-i\ve{k}\cdot \ve{r}}-\rho),
\end{equation}
where $\varrho(\ve{k})$ is the Fourier transform of the object's mass density. If we expand the exponentials in \eqref{eq.reducedCSLmaster} to first order and the center of mass oscillations are taken to be much smaller than $r_c$, we can simplify the CSL diffusion operator~\cite{Nimmrichter2014} to
\begin{align}
\mathcal{L}_{\rm csl}\rho &=-D_{\rm csl}[x,[x,\rho]]/\hbar^2=-\alpha\lambda_{\rm csl}\left(\frac{\hbar}{r_c}\right)^2[\hat x,[\hat x,\rho]]/\hbar^2
\end{align}
where $\alpha$ is a geometric factor describing the shape and density of the object, which for a sphere (following again Ref.~\cite{Nimmrichter2014}) is given by
\begin{equation}
\alpha=\left(\frac{m}{m_0}\right)^2\left[e^{-R/r_c^2}-1+\frac{R^2}{2 r_c^2}(e^{-R^2/r_c^2}+1)\right]\frac{6 r_c^6}{R^6}.
\end{equation} 
We can now incorporate $D_{\rm csl}$ into the master equation describing a macroscopic oscillator in the same way as any other diffusive noise source. 

\subsection*{Dissipation and Diffusion in the CSL Mechanism}
Most noise sources will cause momentum diffusion and momentum dissipation. Typical Brownian noise such as background gas collisions and blackbody radiation take this form, resulting in energy absorption (at rate $D_{\rm diff}$) from the environment surrounding the nanosphere, as well as deposition of excitations into it (contributing to momentum damping at rate $\Gamma$). Other noise sources consist of \emph{one way} mechanisms that result in unidirectional feeding of energy to the nanosphere (thus contributing to the value of $D_{\rm diff}$) and thus heat it up without damping its motion. A relevant source of this form of noise would be the scattering of photons from the cavity field mode~\cite{Romero-Isart2011b}: the oscillator can scatter these photons out of the cavity and gain energy without giving up anything to create them. 

CSL, as a developing theory, has been confronted with various problems, the question of energy conservation amongst them. As we've stated, the mechanism of localization in CSL arises via a coupling to a noise field. Through this constant one-way coupling energy is transferred, and ultimately violates energy conservation. This issue was recently addressed in \cite{Smirne2014a}, in which a new modification to the Schr\"{o}dinger equation is introduced, resulting in a dissipative effect which remedies this problem and restores energy conservation. The rate of dissipation for a nanosphere of $R=100$nm is found to be $\sim 10^{-3}$ when using the Adler value for $\lambda_{\rm csl}$, which is negligible for our considerations, and hence we leave it out of our analysis. As such, we treat noise from collapse as a solely contributing to momentum diffusion, and not to dissipation -- making it analogous to optical scattering. 

%\bibitem{note} In order to conserve energy, CSL must include a dissipative mechanism that would compensate for the energy imparted by random collapses. A. Smirne and A. Bassi, arXiv:quant-ph/1408.6446, have developed a model for such mechanism. As we show in the Supplementary Information, the effect of this mechanism upon our system is negligible.

\section{Conventional Noise Sources}
	
\subsection*{Collisions With the Background Gas}
Collisions with background gas are typically the dominant source of noise for trapped nanospheres (following optical scattering). Since the nanosphere can both give and receive momentum to and from the background gas, it contributes both to $D_{\rm diff}$ and $\Gamma$. The contribution to momentum diffusion is given by 
\begin{equation}
D_{\rm gas}=\frac{\gamma_g k_B T_{\rm env}}{\hbar \omega_m},
\end{equation}
where $T_{\rm env}$ is the temperature of the environment and $\gamma_g$ is given in Ref.~\cite{Chang2010} as $\gamma_g=\frac{16 P}{\pi v_g R d}$. Here, $v_g$ is the mean velocity of the gas particles, $d$ is the density of the sphere and $P$ is the background gas pressure. The contribution to momentum dissipation is simply $\Gamma_{\rm gas}={\gamma_g}/{4}$.

\subsection*{Blackbody Radiation}
Blackbody radiation also contributes to both diffusion and dissipation, being a coupling to a bath with infinite modes at the temperature of the environment. The damping coefficients given in Ref.~\cite{Chang2010} are
\begin{equation}
\gamma_{{\rm bb},i}=\frac{2 \pi^4}{63}\frac{(k_B T_i)^6}{c^5\hbar d \omega_m}\textrm{Im}\frac{\epsilon-1}{\epsilon+2}
\end{equation}
where $i$ represents either emissive or absorptive processes. We have $T_i=T_{\rm env}$ for absorptive radiation and $T_i=T_{\rm int}$ for emissive. This gives us the diffusive term $D_{\rm bb}= {(\gamma_{\rm bb,e} T_{\rm int}+\gamma_{\rm bb,a}T_{\rm env})k_B}/({2 \hbar \omega_m})$ and the dissipative one $\Gamma_{\rm bb}=({\gamma_{\rm bb,e}+\gamma_{\rm bb,a}})/{4}$.

We find that emissive radiation has a far stronger effect than absorptive. We can characterise the internal temperature which is primarily determined by the intensity of the cooling laser. The expression for the steady-state bulk temperature is given in Ref.~\cite{Romero-Isart2010} as
\begin{equation}
T_{\rm bulk}=\left(I_0\frac{4 \pi^3 R}{e \sigma \lambda_{\rm laser}}\frac{3 \epsilon_2}{(\epsilon_1+2)^2+\epsilon_2^2}+T_{\rm env}^4\right)^{1/4}
\end{equation}
where $I_0$ is the intensity of the cooling laser at its waist and $\lambda_{\rm laser}$ is its wavelength, $e$ is the emissivity of sphere,  $\epsilon=\epsilon_1+i\epsilon_2$ is its complex dielectric constant, $\sigma$ is the Stefan-Boltzman constant. 

\subsection*{Electric Field Noise}
There are three mechanisms which will contribute to noise in the electric field experienced by the nanoparticle: Johnson noise, caused by thermal fluctuations in the electrodes; patch potential noise, thought to be caused by adatoms \cite{Safavi-Naini2011} on the surface of the electrodes; and acoustic vibrations in the lab, which will shake the electrodes relative to the trapped particle. Though the origins of these forms of noise are fairly independent, and their spectra will be different, all three will contribute to the heating of the particle and in practice it will be difficult to distinguish between them. They can all be modelled as a variation in the effective spring constant of the trap \cite{Gehm1998}. In ion traps, electric field noise is often dominant over other forms of heating \cite{Eltony2014,Deslauriers2006}. The heating rate due to patch fluctuations is given by 
\begin{equation}\label{eq.PatchHeating}
\Gamma_{\rm patch}=\frac{q^2}{4 m \hbar \omega}S_{E_k}(\omega_m)
\end{equation}
where $S_{E_k}(\omega)=\Lambda(\vec{r})R(\omega)$ is the power spectrum of the electric field noise, determined a geometry factor $\Lambda (\vec{r})$ and the noise spectrum $R(\omega)$. A discussion of $\Lambda(\vec{r})$ can be found in \cite{Low2011}. Whilst the details vary significantly between electrode architectures, a trend is shared in that $\Lambda(\vec{r})\propto d^{-a}$, where $a\approx 4$. We can describe the noise spectrum of the adatoms as \cite{Safavi-Naini2011} $R(\omega)=\sigma S_\mu (\omega)$, where $\sigma$ gives the density of adatoms on a surface and $S_\mu$ is the spectrum of the fluctuating dipole of a single adatom. By utilising the framework of \cite{Low2011} to describe the geometry of a trap architecture, and that of \cite{Safavi-Naini2011} to describe the patch potentials which form on the electrode surfaces, a detailed picture can be built of the noise one would expect from any reasonable shape of trap, made of any metal. However, since we are not proposing a specific electrode architecture, it makes more sense to translate experimentally verified heating rates from other experiments into our own setting. Ref.~\cite{Eltony2014} finds heating rates $D'_{\rm E field}\sim10$ Hz for a single ion with a mechanical frequency $\omega'_m$ in the range of MHz. Assuming a similar trap geometry, we can see via Eq.~\eqref{eq.PatchHeating} that we would expect a heating rate of $D_{\rm Efield}=D'_{\rm Efield} \frac{q^2 m' \omega'_m}{q'^2 m \omega_m}\approx10^{-4}$, which is negligible compared to the other noise sources that have been identified. Note that this prediction is based on experimental data gathered in ~\cite{Eltony2014}, and as such must reflect the presence of all three forms of electric field noise. 

\subsection*{Anisotropy of the Sphere}
The question of sphere anisotropy is a recognized problem when considering the dynamics of a trapped nanosphere, and is addressed in the supplementary information of ~\cite{Chang2010}. In their analysis, the anisotropy presents a problem because it will cause an increased polarizability. This problem is resolved in our case, since there will be no optical field over the period of interest. The potential problem arising from anisotropy for us will be the corresponding anisotropy of the charge distributed over the surface, though we remark that this distribution may also be anisotropic even for an ideal sphere. This question of charge distribution is addressed below.  
 
\subsection*{Rotational Dynamics}
The anisotropy of the charge distribution on the sphere will cause a dipole moment, and this dipole moment will experience a torque due to the electric fields, and further it will affect the COM motion. The equations of motion for the COM are given by 
\begin{equation}
m {\bf \ddot{r}}-q\, {\bf E(}{\bf r},t)-\nabla [ \boldsymbol{\mu}\cdot {\bf E}({\bf r},t)]=0
\end{equation}
where ${\bf r}=(x,y,z)$ is the position of the nanosphere in three dimensions, $\boldsymbol{\mu}$ is its dipole moment, and $\bf{E}({\bf r},$$t)$ is the electric field at point ${\bf r}$ at time $t$. Clearly we can see a coupling between the orientation of the dipole moment and the COM motion. This could potentially constitute a heating source. In order to understand what this heating rate amounts to, we need to model the rotational dynamics of the dipole together with the translational motion of the sphere, and see what the effect of the dipole moment is upon the COM motion. 

This situation is very similar to that of diatomic molecules held in equivalent traps, and as such we can draw upon the relevant literature. In modelling the rotational dynamics for systems of this type, it is a well known problem that the equations of motion cannot be solved using spherical polar coordinates ~\cite{Hashemloo2015}. Working with the quaternion formalism~\cite{Evans1977,Evans1977}, we built a numerical simulation in XMDS \cite{Dennis2013} which modelled the coupled dynamics of the rotary and COM motion, and experimented with the effects of different dipole moment magnitudes. We expect that packing the maximum possible charge onto the surface of the sphere would naturally lead to the most isotropic distribution ~\cite{Hinds1999}, which in our case gives us $q=1500$ eV. We find that the effect of the rotary motion does cause an appreciable heating effect for a large enough dipole moment, which occurs when $\mu_0= \,R \times 5 \,{\rm eV}$. This limit on the dipole magnitude $\mu_0$ is perhaps more clearly expressed as a ratio ${q_{\rm edge}}/{q_{\rm core}}\leq {1}/{300}$, where we have simplified the charge distribution into $q=q_{\rm core}+q_{\rm edge}$, $q_{\rm core}$ being the charge which is completely isotropic and effectively bunched at the COM, and $q_{\rm edge}$ being the charge which is bunched at one position of the surface, and hence causes the dipole moment. 

We can infer a bound on the ratio $\frac{q_{\rm edge}}{q_{\rm core}}$ from experimental data already gathered. As stated, this experimental proposal rests upon the hybrid type trap which has already been constructed and used. In the experiments which we have conducted so far, we have used far lower charge numbers, which would presumably distribute themselves more anisotropically the situation presented in this paper. In analysing the dynamics of these lower-charge systems we do not find strong evidence of the coupling between the rotational motion to the translational dynamics of the sphere. We take this as an indication that in these situations we indeed have $\frac{q_{\rm edge}}{q_{\rm core}}\leq \frac{1}{300}$. Therefore, we conclude that in our scenario, in which we would presumably have a more isotropic distribution, we would have an even lower ratio and therefore an acceptably low heating rate from the rotational dynamics.

%The potential for the ion trap is given by $\frac{V_0}{r_0^2}(x^2+y^2-2 z ^2)\cos(\omega_d t)$ where $V_0$

%\bibliography{library.bib,ManualLibrary.bib}

\end{document}